\documentclass[numbers,webpdf,imaiai]{ima-authoring-template}%

\graphicspath{{Fig/}}

\usepackage[utf8]{inputenc}
\usepackage[T1]{fontenc}
\newcommand{\eq}[2]{\begin{equation} \label{eq:#1} #2 \end{equation}}
\newcommand{\dmath}[1]{\begin{displaymath} #1 \end{displaymath}}
\usepackage{calrsfs}
\DeclareMathAlphabet{\pazocal}{OMS}{zplm}{m}{n}
\DeclareMathAlphabet{\mathbfit}{OML}{cmm}{b}{it}
\usepackage{graphicx}
\newcommand{\hilbert}{\pazocal{H}}
\usepackage{braket}
\usepackage{musicography}
\usepackage{indentfirst}
\theoremstyle{thmstyletwo}%
\numberwithin{equation}{section}

\begin{document}

%\DOI{DOI HERE}
%\copyrightyear{2021}
%\vol{00}
%\pubyear{2023}
%\access{Advance Access Publication Date: Day Month Year}
%\appnotes{Paper}
%\copyrightstatement{Published by Kazuki Otsuka. All rights reserved.}
%\firstpage{1}

%\DOI{}
\copyrightyear{2023}
%\vol{}
\pubyear{2023}
%\access{}
%\appnotes{}
\copyrightstatement{Published by Kazuki Otsuka. All rights reserved.}
%\firstpage{1}

%\subtitle{Subject Section}

\title[]{Computational Language $\beta$ based on Orthomodular Lattices with the Non-distributivity of Quantum Logic}

\author{Kazuki Otsuka
\footnote{
  %The Graduate School of Interdisciplinary Information Studies, The University of Tokyo, 7-3-1 Hongo, Bunkyo-ku, 113-0033, Tokyo, Japan. 
  This paper was written and published as a master's thesis (Jan 2023).
  %\ Email:otsuka.kazuki@googlemail.com
}
\\
\address{
  \orgdiv{The Graduate School of Interdisciplinary Information Studies},
  \orgname{The University of Tokyo},
  \orgaddress{
    \street{7-3-1 Hongo, Bunkyo-ku},
    \postcode{113-0033},
    \state{Tokyo},
    \country{Japan}
    }
  }
}
%\address{
%  \orgdiv{},
%  \orgname{},
%  \orgaddress{
%    \street{},
%    \postcode{},
%    \state{},
%    \country{}
%    }
%  }
%}

%\authormark{K.Otsuka}

%\corresp[*]{Corresponding author: \href{email:email-id.com}{email-id.com}}

%\received{Date}{0}{Year}
%\revised{Date}{0}{Year}
%\accepted{Date}{0}{Year}

%\editor{Associate Editor: Name}
%\vspace{1cm}
\abstract{
  \begin{center}
    \textbf{Abstract}
  \end{center}
\indent It is argued that transformation processes (generation rules) showing evidence of a long evolutionary history 
in universal computing systems can be generalized. 
The explicit function class $ \Omega $ is defined as follows: 
`Operators whose eigenvectors (or eigenvalues) have an irrational number in their components constitute a class of functions with quasi-periodic structure,  
$ \Omega $, and the class $ \Omega $ shows evidence of a long evolutionary history.'
In order to empirically proove this theorem by examining physical systems carrying out life activities or 
intellectual outputs of developed intelligence, 
the basic framework of the universal machine model $ \pazocal{C} $ 
and the computational language $ \beta $ is presented as a model for general computational methods,
which allow transformation processes (generation rules) with deep algorithmic complexity 
to be derived from generation results.
$ \pazocal{C} $ and $ \beta $ perform massively parallel computations on event-state systems 
consisting of exponential combinations of propositional elements expressed 
in terms of correlations between subsystems.
The logical structure of the computational language relies on a non-distributivity in Hilbert spaces 
or orthogonal modular lattices, 
allowing for the manipulation and deduction of simultaneous propositions.
In this logical local structure, the propositions implying certain consequences are not uniquely determined.
}
%\keywords{keyword1; Keyword2; Keyword3; Keyword4.}

% \boxedtext{
% \begin{itemize}
% \item Key boxed text here.
% \item Key boxed text here.
% \item Key boxed text here.
% \end{itemize}}

\maketitle

%%% Chapter 1
\vspace{1cm}
\section{Why Beautiful Things are Beautiful}
%The introduction introduces the context and summarizes the manuscript. It is importantly to clearly state the contributions
%of this piece of work. 

When a sequence of symbols is generated with a high degree of complexity, 
there is no general method for deriving the transformation process from the generated results.
Supposing there is a sequence of binary symbols as follows:

\eq{1.1}{
0101010101010101010101010101010101
}
\eq{1.2}{
111110000111001110011000111100000
}
\eq{1.3}{
10101 1010 1010 101 10 10 101 1010 10101
}
The first one is an alternating sequence of 0 and 1 and appears to be the simplest of the three sequences. 
The second is symmetric if one looks carefully. 
The degree of complexity of the first and the second are both small, but the second seems to be slightly more complex.
The third is a combination of the alternation rules of the first and the symmetry rules of the second and 
can be said to contain the complexity of the generative rules of the first and the second. \\

Computational complexity theory has defined the complexity of generation processes 
as the computing time of programs that efficiently give rise to a given 
$ \mathbb{Z} \rightarrow \mathbb{Z} $ on a Turing-equivalent system. 
Most systems with computational universality generate computable functions for 
$ \mathbb{Z} \rightarrow \mathbb{Z} $ by applying basic finite procedures recursively.
This complexity corresponds to the amount of evolution required to evolve from a precursor in the evolution of life, 
and similarly to the amount of evolution required to create new knowledge from existing intellectual output
in terms of the growth of human knowledge. 
Therefore, it is reasonable to assume that the "evidence" imprinting the long "causal history" 
that arose in the natural world under a "certain order" is preserved in some form or encoded information.
From the principle of natural selection, 
it is also reasonable to think that life acquires the ability to intuitively discern this degree of complexity 
through the generated results, 
and that this ability enables it to distinguish between more evolved and less evolved forms.

On the other hand, it is not generally possible to distinguish from the results generated by a universal computing system 
whether those results are the output of a complex transformation process or the result of a random behavior.

\eq{1.4}{
110101100011010101100010101010010110101
}
\eq{1.4}{
0 1 0 0 1 010 01001 01001 01001010 0100101001010
}
The first of these two binary sequences is a random output, 
while the second is the result of applying the Fibonacci word generation rules nine times.
In principle, one cannot rule out the possibility that 
a randomly generated result may coincide with a result generated by a complex transformation process. 
The second one can be generated with a probability of 
$ \frac{1}{2^{34}} $
. 
It is easy to infer that 
the application of consistent rules to systems of a certain size is required 
in order to show evidence that the morphology of the generated results is due to a complex evolutionary process.

It is known that a pair of natural numbers $ (p, q) $ generated by 
the Fibonacci rule maximizes the computational complexity of the Euclidean algorithm 
for finding the greatest common divisor (gcd). 
That is, if the quotient of $u$ divided by $v$ is $s$ and the remainder is $r$, then for 
\eq{}{
  \begin{array}{c}
    u_{(1)} = v_{(1)}s_{(1)}+r_{(1)}  \\
    u_{(2)} = v_{(2)}s_{(2)}+r_{(2)}  \\
    \ \ \ \ \ \ \ \  \vdots \\
    u_{(t)} = v_{(t)}s_{(t)}+r_{(t)} \\
    \\
     %(p,\ q,\ r,\ t \in \mathbb{N},\ u_{(1)} = q,\ v_{(1)} = p) \\
     %(u_{(1)} = q,\ v_{(1)} = p,\ \ p,\ q,\ r,\ t \in \mathbb{N}) \\
    \end{array}
}
where $ u_{(1)} = q,\ v_{(1)} = p,\ \ (p,\ q,\ r,\ t \in \mathbb{N})$, the following is obtained
\dmath{ 
  %\text{for\ all\ } t,\ s_{(t)} = 1,\ u_{(t+1)} = v_{(t)},\ v_{(t+1)} = r_{(t)}
  \forall t,\ s_{(t)} = 1,\ u_{(t+1)} = v_{(t)},\ v_{(t+1)} = r_{(t)}
}
and the Fibonacci process that generated $q$ (and $p$) is exactly reversed until $p=1$ and $q=2$.
This fact suggests that transformations by the Fibonacci rule provide a way 
to show evidence of a long evolutionary history of causality in nature. 
That is, the results of multiple applications of a transformation rule are recorded 
in the form they are as proof of the generation costs.

Fibonacci numbers are generated by 
using the operator 
\eq{fibmatrix}{ 
  \pazocal{F} = 
  \begin{bmatrix} 
    1 & 1 \\ 
    1 &  
  \end{bmatrix}
}
with a recursive process
\eq{}{
  \begin{array}{l}
    \mathbfit{u}_{t+1} = \pazocal{F} \mathbfit{u}_{t}, \ \  
    \mathbfit{u}_{0} = 
      \begin{bmatrix} 1 \\ 0 \end{bmatrix} 
  \end{array}
}
The matrix $ \pazocal{F} $ is a self-adjunct Hermitian operator with real eigenvalues including 
$ \tau = \frac{ 1 + \sqrt{5} }{2}$.
%and $ \bar{\tau} = \frac{ 1 - \sqrt{5} }{2}$. 
It is also invertible and the transformation 
$ T : G \rightarrow G $ 
is bijective and hence logically reversible. 
Its spectral decomposition is a linear combination of the projection operators on the eigenvectors containing $ \tau $.

Let the eigenvalues of $ \pazocal{F} $ be 
$ \lambda_{1} = \frac{1+\sqrt{5}}{2} (= \tau ) 	\approx 1.618 $,
$ \lambda_{2} = \frac{1-\sqrt{5}}{2} \approx-.618 $
and the eigenvectors  
\eq{}{ 
  \mathbfit{x}_{1} = 
    \begin{bmatrix} 
      \lambda_{1} \\ 
      1  
  \end{bmatrix}, \ \ \
  \mathbfit{x}_{2} = 
    \begin{bmatrix} 
      \lambda_{2} \\ 
      1
    \end{bmatrix}
}
then the diagonalization of $ \pazocal{F} $ 
%\eq{}{
%  \pazocal{F} = 
%    \begin{bmatrix} 
%                     &                  \\
%    \mathbfit{x}_{1} & \mathbfit{x}_{2} \\
%                     &                  
%    \end{bmatrix}
%    \begin{bmatrix} 
%    \lambda_{1} &              \\
%                & \lambda_{2}  \\
%    \end{bmatrix}
%    \begin{bmatrix} 
%        \frac{1}{\sqrt{5}} & - \frac{ 1 - \sqrt{5} }{ 2 \sqrt{5} } \\
%      - \frac{1}{\sqrt{5}} &   \frac{ 1 + \sqrt{5} }{ 2 \sqrt{5} }
%    \end{bmatrix}
%}
gives
%\eq{}{
% \pazocal{F} \mathbfit{u}(0) =
%   \begin{bmatrix}
%                      &                  \\
%     \mathbfit{x}_{1} & \mathbfit{x}_{2} \\
%                      &                  
%   \end{bmatrix}
%   \begin{bmatrix} 
%   \lambda_{1} &              \\
%               &  \lambda_{2} 
%   \end{bmatrix}
%   \frac{1}{\lambda_{1} - \lambda_{2}}
%   \begin{bmatrix} 
%     1   \\
%     -1  
%   \end{bmatrix}
%}
%.
%Hence, 
\eq{}{
  \pazocal{F}^{n} u(0) = 
    \frac{1}{\lambda_{1} - \lambda_{2}}
    \begin{bmatrix}
                 &                  \\
      \mathbfit{x}_{1} & \mathbfit{x}_{2} \\
                 &                  \\
    \end{bmatrix}
    \begin{bmatrix}
      \lambda_{1}^{n} &                \\
                      & \lambda_{2}^{n}
    \end{bmatrix}
    \begin{bmatrix}
      1 \\
      -1
    \end{bmatrix}
%    =
%    \begin{bmatrix}
%                 &                  \\
%      \mathbfit{x}_{1} & \mathbfit{x}_{2} \\
%                 &                  \\
%    \end{bmatrix}
%    \begin{bmatrix}
%      \frac{ \lambda_{1}^{n} }{ \lambda_{1} - \lambda_{2} }   \\
%      \frac{ \lambda_{2}^{n} }{ \lambda_{1} - \lambda_{2} }
%    \end{bmatrix}
%%    =
%%    \frac{1}{ \lambda_{1} - \lambda_{2} }
%%    \begin{bmatrix}
%%      \lambda_{1}^{n}  \mathbfit{x}_{1}  \\
%%      \\
%%      \lambda_{2}^{n}  \mathbfit{x}_{2} 
%%    \end{bmatrix}
    =
    \frac{1}{ \lambda_{1} - \lambda_{2} }
    \begin{bmatrix}
      \lambda_{1}^{n+1} - \lambda_{2}^{n+1} \\ 
      \\
      \lambda_{1}^{n} - \lambda_{2}^{n} 
    \end{bmatrix}
}
This equation shows that, the number generated by applying the Fibonacci rule $ n $ times is
\eq{}{
  \phi(n)= \frac{ \lambda_{1}^{n+1} - \lambda_{2}^{n+1} }{ \lambda_{1} - \lambda_{2} }
}
since $ \lambda_{2} < 0 $, 
the ratio 
%\eq{}{
%   \phi(n+1)/\phi(n) = r  
%}
%is 
%\eq{}{
%  \lim_{n \rightarrow \infty} r \rightarrow \lambda_{1} (= \tau)
%}
converges 
\eq{}{
  \lim_{n \rightarrow \infty} \phi(n+1)/\phi(n) \rightarrow \lambda_{1} (= \tau)
}
% because of 
% $ \lambda_{2} < 0 $
it means multiple applications of the Fibonacci process bring the ratio closer to $\tau$.
This $ \tau $ is generally called the golden ratio.
Now, note that $\tau$ contains an irrational number $\sqrt{5}$.
From the polar decomposition 
\eq{}{
  \pazocal{F} = U \Sigma V^{T} = (UV^{T})(V \Sigma V^{T}) = (Q)(S)
}
of $ \pazocal{F} $, 
the components of the rotation matrix $Q$ are
\eq{}{
  \mathbfit{u}_{1} = \frac{\mathbfit{x}_{1}}{|\mathbfit{x}_{1}|},\  
  \mathbfit{u}_{2} = \frac{\mathbfit{x}_{2}}{|\mathbfit{x}_{2}|},\
  \mathbfit{v}_{1} = \frac{A^{T}\mathbfit{x}_{1}}{\sqrt{\lambda_{1}}|\mathbfit{x}_{1}|},\  
  \mathbfit{v}_{2} = \frac{A^{T}\mathbfit{x}_{2}}{\sqrt{\lambda_{2}}|\mathbfit{x}_{2}|},\
}
that each consist of the left singular vector matrix $U$ and the right singular vector matrix $V^{T}$. 
Expanding this $Q$, the upper left component is 
$ \cos \theta =  \frac{ 4+ \sqrt{5} }{ \sqrt{2} \sqrt{ 35-15 \sqrt{5} } } +  \frac{ -1 + -\sqrt{5} }{ \sqrt{2} \sqrt{ 30+15 \sqrt{5} } } $
, which still contains an irrational number.
In general, 
\eq{}{
 \cos \theta = 
   \frac{ a_{11} x_{11}^{2} + a_{12} x_{11} x_{12} }{ \sqrt{\lambda_{1}} |\mathbfit{x}_{1}|^{2} } + 
   \frac{ a_{11} x_{21}^{2} + a_{12} x_{21} x_{22} }{ \sqrt{\lambda_{2}} |\mathbfit{x}_{2}|^{2} }
}
where $a_{ij}$ is the $(i, j)$ component of the original matrix.
According to the definition by Bohr (1946), 
if there exists an "\emph{almost parallel shift}" $t$ satisfying 
\eq{}{
  \forall x \in \mathbb{R},\ \ \epsilon > 0, \ \ \ | f(x+t)-f(x) | < \epsilon \ \
}
$f(x)$ is said to be an "\emph{almost periodic}" function. 
And $ f(x) = \sin(x) + sin(\tau x) $ has this quasi-periodic pattern, 
i.e., an ordered but never repeated structure.

Assuming that transformations by the Fibonacci rule are evidence of a long evolutionary history of causality, 
the results produced by such transformations 
have the structure of an asymmetric transformation that has a discernible basic pattern, 
but, at the same time, do not match no matter how many times the transformation is repeated.
On the other hand, the symmetry group $S_{3} : G \rightarrow G$ can be described by six operations 
$ \{1,\ x,\ x^2,\ y,\ xy,\ x^2y \} $,\ %, where $ x: \{a,\ b,\ c \} \rightarrow \{c,\ a,\ b \},\ y: \{a,\ b,\ c \} \rightarrow \{b,\ a,\ c \}$,
and the result always coincides with one of the rotations represented by 
$ e^{ \frac{ n }{ 6 } i 2 \pi } \ \ ( n \in \{ 1, \dots, 6 \} ) $
. 
In contrast to quasi-periodic transformations, 
symmetric transformations do not increase in complexity in the sense of the example at the beginning, 
or only in combination with quasi-periodic transformations. 
Therefore, they do not by themselves provide direct evidence of a complex evolutionary process. 

Thus, these arguments can be generalized and I propose
%\textit{Theorem\ 1}
%\eq{}{
%  \begin{array}{l}
%    \text{
%      \textit{Theorem\ 1}
%    } \\
%    \ \ \ \text{
%    \textit{ 'Operators\ whose\ eigenvectors have irrational numbers in their components constitute }
%    } \\      
%    \text{ 
%    \textit{\ \ \ \ a class of functions with quasi-periodic structure,  $ \Omega $ '}
%    }
%  \end{array}
%}
\begin{flalign}
    & \ \ \ \  \text{
      \textit{Theorem\ 1}
    } \nonumber&\\
    &   \ \ \ \ \ \ \ \ \ \ \ \ \ \ \ \ \ \ \ \ \text{
    \textit{ `Operators\ whose\ eigenvectors\ (or\ eigenvalues) have an irrational number in their }
    } \nonumber&\\      
    & \text{ 
    \ \  \ \ \ \ \ \ \ \ \ \ \ \ \ \ \ \ \textit{\ \ \ \ components constitute a class of functions with quasi-periodic structure,  $ \Omega $ '}
    } &
\end{flalign}
\vspace{-0.86cm}
\begin{flalign}
  & \ \ \ \ \text{
      \textit{Therem\ 2} \nonumber
    }  &\\
   &  \ \ \ \ \ \ \ \ \ \ \ \ \ \ \ \ \ \ \ \ \ \text{ 
    \textit{ `The class of functions $ \Omega $ shows evidence of a long evolutionary history.' }
    }  &
\end{flalign}
%\textit{Therem\ 2}
%\eq{}{
%  \begin{array}{l}
%    \text{
%      \textit{Therem\ 2}
%    } \\
%    \ \ \ \text{
%    \textit{ 'The class of functions $ \Omega $ shows evidence of evolutionary history.' }
%    } \\      
%  \end{array}
%}
Then I add a new interpretation of the \emph{golden ratio}.
\begin{flalign}
    & \ \ \ \  \text{
      \textit{Theorem\ 3}
    } \nonumber&\\
    &   \ \ \ \ \ \ \ \ \ \ \ \ \ \ \ \ \ \ \ \ \text{
    \textit{ `The golden ratio is a ratio at which the long history of evolution by a function}
    } \nonumber&\\      
    & \text{ 
    \ \  \ \ \ \ \ \ \ \ \ \ \ \ \ \ \ \ \textit{\ \ \ \  in the function class $\Omega$ converges.'}
    } &
\end{flalign}

There are a total of $16$ $ 2\times2 $ matrices made by binary $1$ and $0$, 
of which only $\pazocal{F}$ satisfies the condition (1.17).
%As an example of these binary operators
%\eq{}{
%  \begin{bmatrix}
%    1 & 0 \\ 
%    0 & 0
%  \end{bmatrix} \ \ \  \text{\textit{Elimination}} \ \ \
%  \begin{bmatrix}
%    1 & 0 \\ 
%    1 & 0
%  \end{bmatrix} \ \ \  \text{\textit{Replication}} \ \ \
%  \begin{bmatrix}
%    1 & 0 \\ 
%    0 & 1
%  \end{bmatrix} \ \ \  \text{\textit{Identity}} \ \ \
%  \begin{bmatrix}
%    0 & 1 \\ 
%    1 & 0
%  \end{bmatrix} \ \ \  \text{\textit{Permutation}} \ \ \
%  %[1 0 \\ 0 0 ] Elimination \\
%  %[1 0 \\ 1 0] Duplication \\
%  %[1 0 \\ 0 1] Constant \\
%  %[0 1 \\ 1 0] Replace \\
%}
We can consider a class of primitive recursive functions with these binary linear operators as basic functions,
and this issue is then discussed below.

The paradigm of non-reductionistic systems in which the parts and the whole are interdependent was presented by Anderson (1972), 
where a new order arises that did not exist in the subsystems alone. 
Cellular automata with computational universality represent the behavior of finite interdependent systems 
and provide a model for studying primitive recursive functions with simple binary rules.
For example, Cook (2004) proved that a 1-dimensional cellular automata,
where a cell becomes 1 if a right neighbor is 1 (rule 1) and becomes 0 if both neighbors are 1 (rule 2), 
can emulate the Post normal canonical system and then is Turing-complete.
\dmath{
  \begin{matrix}
    0 & 0 & \underline{1}
  \end{matrix} \ \ 
  \rightarrow \ \ 
  \begin{matrix}
    0 & \bold{\underline{1}} & 1
  \end{matrix} \ \ 
  \rightarrow \ \ 
  \begin{matrix}
    \bold{\underline{1}} & 1 & \underline{1}
  \end{matrix} \ \ 
  \rightarrow \ \ 
  \begin{matrix}
    1 & \bold{0} & 1
  \end{matrix}
}
It can be proved elementary that there is no finite binary linear operator that expresses this production rule.
For example, a linear transformation representing (rule 1)
\eq{}{
  \begin{bmatrix}
    1 & 1 &   \\
      & 1 & 1 \\
      &   & 1
  \end{bmatrix}
}
works correctly on inputs \
$\begin{matrix}0 & 0 & 1\end{matrix}$, \ \ \ \ 
$\begin{matrix}0 & 1 & 0\end{matrix}$, \ \ \ \ 
$\begin{matrix}1 & 0 & 0\end{matrix}$, \ \ 
and\ \ 
$\begin{matrix}1 & 0 & 1\end{matrix}$, 
but $2$ is generated from inputs 
$\begin{matrix}1 & 1 & 0\end{matrix}$, 
and 
$\begin{matrix}0 & 1 & 1\end{matrix}$
. 
A complete representation does not exist under the classical algebraic group laws.

On the other hand, 
if it is assumed that this output "value" simply gives a probability distribution 
over a finite number of binary rules of the system, 
it may be possible to claim that the operator correctly describes the behavior. 
But this requires a change in logic or some different framework.
One possibility would have to assume that there is not just one proposition $\pazocal{P}$ (cause) 
that implies a proposition $\pazocal{S}$ (generated result).

Just as life was once thought to be a creation of God, 
but not many people believe that now, 
so the idea that all mathematical logic is not the "language of God" has become more common, 
especially since the establishment of non-Euclidean geometry.
With the birth of quantum theory at the beginning of the 20th century, 
it became clear that we live in a world with a non-classical logic, 
while classical logic is organized on a very specific lattice, the Boolean lattice.

According to Putnam (1968), unlike classical logic where 
\begin{quote}
	\textit{"there was (idealizing) one proposition $\pazocal{P}$ 
  which was true of $\pazocal{S}$ and such that every physical proposition about 
  $\pazocal{S}$ was implied by $\pazocal{P}$" }
\end{quote}
in the structure of non-distributive quantum logic on orthogonal modular lattices, 
$ (S_{1} \land S_{2} \land \cdots \land S_{R} ) $ can be true even if 
$ S_{1},\ S_{2}, \ldots ,\ S_{R} $ are all false.
Therefore, if T1 is true, it is possible for 
\eq{put1}{
 T_{1} \cdot (S_{1} \lor S_{2} \lor \cdots \lor S_{R})
}
to be true and simultaneously 
\eq{put2}{
 T_{1} \cdot S_{1} \lor T_{1} \cdot S_{2} \lor \cdots \lor T_{1} \cdot S_{R} 
}
to be false (failure of the distribution law). 

These two propositions (\ref{eq:put1}) and (\ref{eq:put2}), 
which were equivalent in classical logic, 
are mapped into two different subspaces on the space $ \mathbb{L}(\hilbert) $ 
representing all propositions of the system. 
The theory was formulated by Birkhoff and Neumann (1936) 
as a system where arbitrary observable $ A $ is described by self-adjoint operators. 
In this projection logic on Hilbert spaces, 
the mapping of propositions represented by state vectors to closed subspaces expresses (probabilistic) truth or falsehood.

After a few decades, the universal machine model of quantum computers was then formulated by Deutsch (1985) for the first time, 
but since then it has been studied mainly for the complete simulation of the time evolution of physical systems, 
and there are not many theoretical models of computational languages that 
explicitly use the non-classical logic structure of quantum logic.
Including QTM, Many Turing-equivalent systems exist.
Shannon (1956) posed the problem of a minimal Turing machine, 
which was later proved by Minsky (1961) to require two tapes and three variables.
On the other hand, the logic or language system 
that can be constructed on a universal computer remains a challenge for future research.

I call a transformation that expresses rules that provide evidence of deep evolution, 
"\emph{universal algorithms}".
These rules include a class of quasi-periodic transformations $ \Omega $.
It seems reasonable to rely on computations that assume qubits with the "genuine" microscopic logical structure 
of nature to verify the existence of universal algorithms and to decode them. 
This is because, in order to verify the "genuine" transformation rule, 
it is reasonable to choose the natural empirical logic for obtaining the class of basic functions 
that make up a general recursive function.

"\emph{Universal}", in the sense of computability presented by Deutsch (1985), 
i.e., a computation in which the physical system is representable, 
refers to a transformation algorithm that exists universally (nonlocally) in nature. 
It also means that its generation rules are independent of the decoder. 
Hofstadter (1979) discussed this universality of "meaning" mainly by the Post normal canonical system.
The theoretical models of computers $ \pazocal{C} $ and their language $ \beta $, 
that performs deductive reasoning to derive transformation algorithms from the generated results, 
based on quantum logic are presented in the following chapter.
$ \pazocal{C} $ and $ \beta $ process event-state systems consisting of exponential combinations of propositional elements in a nonclassical logic structure, through massively parallel computation.
In this structure, there is no single proposition implying a certain generated outcome.

On the issue of logical depth, Benett (1994) formulated algorithmic complexity 
as an absolute measure that includes correlations between different parts.
One background is that a few percent of the code that exists between distant parts of the genome of higher organisms 
does not strictly determine the structure of any part, 
but at the same time contains mutual information to convey the relationships between the parts.
The detection of nonlocal correlations, or "\emph{coherence}," 
is essential to distinguish truly profound transformations from coincidentally generated results. 
In short, if a "transformation" occurs similarly in one place as in another, 
then the transformation is justified in being non-accidental. 
When describing the interpretation of physical reality, 
Mermin (1999) argued that correlations are the only objective and reducible units of composition.
In other words, we can say the following. 
There is no absolute causality in a non-element-reductive complex system 
in which the parts and the whole are interdependent, 
and a complete description of the system can only be achieved through correlations between the subsystems.

The computation language $\beta$ and the computer $\pazocal{C}$ are based on the composite system spanned 
by the propositional dimension of $ \pazocal{O}(\prod\ M_{(n)}) $, 
which is exponentially larger than the dimension of the Hilbert space created by each of the subsystems, 
i.e., the sum $ \pazocal{O}(\sum M_{(n)}) $ of the \emph{propositional elements} described in the next chapter. 
By this difference, the local (or nonlocal) correlation among multiple subsystems (propositional elements) is expressed.

In the subsequent Chapter II, 
Section 2.1 first formulates an event-state system 
and a recursive model of the universal computer $ \pazocal{C} $. 
The section also defines the variable types placed 
in the address space for the language $ \beta $, 
including correlated state types 
or combined state types generated by multiple propositional elements. 
Section 2.2 discusses and prooves Turing completeness in $ \pazocal{C} $. 
Section 2.3 presents a mathematical model of a linear logical operator 
that realizes the conditional expressions 
in the non-classical logic structure 
presented at the end of Section 2.1.

For a detailed description of quantum logic in Hilbert spaces 
(or orthomodular lattices) 
and quantum information theory, 
see Cohen (1989) and Peres (1995).

\section{Computational Language $\beta$}\label{sec2}
\subsection{Composite Event-state Systems and the Model of Computers $\pazocal{C}$}

\indent 
Computers $\pazocal{C}$ have an infinite memory space $ \ket{\xi} (i \in \mathbb{N}_{0}) $ 
and a finite bit instruction register $ \ket{\gamma} $ that takes the general form of 
\eq{register}{
  g(\mathbfit{i},\mathbfit{x},\mathbfit{y},\phi(t),k) 
    = \ket{
        \underbrace{ \mathbfit{i} }_{ \text{instructions} }; \
        \underbrace{ x_{1}, \ldots, x_{\bar{M}} }_{ \text{source states} }; \
        \underbrace{ y_{1}, \ldots, y_{\bar{M}} }_{ \text{target states} }; \ 
        \underbrace{ \phi_{1}(t), \ldots, \phi_{\bar{M}}(t) }_{ \text{joint spectra at t} }; \ 
        \underbrace{ k }_{ \text{indices} }
      }
    = \ket{\gamma}
}
where 
$ \mathbb{N}_{0} \in \{0, 1, \ldots \} $, \
$ \mathbb{Z}_{2} \in \{0, 1\} $,  \
$ x_{m},\ y_{m} \in \mathbb{Z}_{2} $, \
$ \phi(t) \in \mathbb{N}_{0}  $, \
$ i,\ k \in \mathbb{N}  $, 
and $ \bar{M} $ is the number of computational bases to be described below.
%$ \mathbb{N}_{N}=\{1, \dotsc , N\} $. 
%\dmath {
%  \ket{\phi}_{m}(t) = \ket{\phi}_{1}(t), \ldots, \ket{\phi}_{\bar{M}}(t) \} \ \ (\phi(t) \in \mathbb{N}_{0} )
%}
%is a finite discrete spectra.
\\
Their computational basis state is a composite system of $ N $ subsystems, 
each with $ M^{(n)} (n \in \mathbb{N}_{N}) $ $2$-state qubits
\eq{2.1}{
\ket{\psi} = \ket{sub\ 1} \otimes \dots \otimes \ket{sub\ N} \in \pazocal{H}^{\bar{M}}\ \ (N \in \mathbb{N})
}
defined on $ \bar{M} $-dimensional Hilbert space $ \hilbert $.
Each subsystem is represented as
\eq{subn}{
  \ket{sub \, n} = \sum_{i=0}^{M^{(n)}-1} \alpha_{i}^{(n)} \ket{i}
}
Each $ M^{(n)} $ is assumed to satisfy the condition $ M^{(n)}=2l^{(n)} (l \in \mathbb{N})$.
Thus the number of standard bases for the composite system $ \ket{\psi} $ is
\eq{2.3}{
\bar{M}=M^{(1)}M^{(2)}\dots M^{(N)}=\prod_{n=1}^{N}M^{(n)}
}
This number is obviously larger than the sum of the number of states each subsystem originally had, 
$ \sum_{n=1}^N M_{(n)} $. As mentioned in Chapter 1, 
this is because "\emph{correlation information}" is contained in this difference, 
and the \emph{correlated states} as a variable type defined by this are described below. 
This correlational information is essential and is the only universal logical foundation 
that encompasses causality as its substructure.
%Correlations are the only universal logical foundation that encompasses causality as its substructure.\\

The set of standard bases for $\pazocal{C}$ is denoted by
\eq{2.4}{
  \pazocal{N} = \{ \ket{m} | m \in \mathbb{N}_{\bar{M}}\}
}
Each element $ \ket{m} $ of $ \pazocal{N} $ corresponds to a state $ s_{m} $ on $ \bar{M} $-dimensional Hilbert space logic $ \mathbb{L}(\hilbert)$, 
i.e., a $1$-dimensional closed subspace of $ \mathbb{L} $. 
$ \mathbb{L}(\hilbert) $ is the semi-ordered set of all subspaces of $ \hilbert \in \mathbb{C}^{\bar{M}} $.
Each state $ s_{m} $ can be regarded as representing a unique logical conjunction of the set of all subsystem states 
\eq{2.5}{
  \Theta = \{ 
    \overbrace{ s_{0}^{(1)},\dots,\ s_{\bar{M}^{(1)}-1}^{(1)}}^{states\ of\ subsystem\ 1},\
    \overbrace{ s_{0}^{(2)},\dots,\ s_{\bar{M}^{(2)}-1}^{(2)}}^{states\ of\ subsystem\ 2},\ 
    \dotsm\ ,\ 
    \overbrace{ s_{0}^{(N)},\dots,\ s_{\bar{M}^{(N)}-1}^{(N)} }^{states\ of\ subsystem\ N}
  \}
}
Elements of $ \Theta $ correspond to $ \ket{i} $ in (\ref{eq:subn}).
Let us refer to these \emph{propositional elements} as the most primitive units of event-state systems.
Examples of propositional elements are a side of a coin, whether a right neighbor cell in a cellular automaton is active or not, 
and whether a note is written as $ E $\musFlat{} or not.

Here a minimal composite system $ \ket{\gamma_{AB}} $ with two distinct 1-qubit subsystems $ \ket{\gamma_{A}}  $ and $ \ket{\gamma_{B}} $ is considered. 
Let $ s_{A} $ and $ s_{B} $ be the corresponding event states (or propositional elements) to $ \ket{\gamma_{A}} $ and $ \ket{\gamma_{B}} $, respectively. 
$ \ket{\gamma_{A}} $ and $ \ket{\gamma_{B}} $ each represent a subspace spanned by two orthogonal $1$-dimensional closed subspaces, 
which can be viewed as units representing the states $ s_{A}\lor s_{A}^{\perp} $ and $  s_{B}\lor s_{B}^{\perp} $ on $ \mathbb{L}(\hilbert) $.
Thus, the composite system $ \ket{\gamma_{A}} \otimes \ket{\gamma_{B}} $ constitutes a $\sigma$-complete Hilbert space logic 
whose states are all $\sigma$-additive sets of $ \{s_{a},\ s_{b} \}$,
and is represented as follows
\eq{2.6}{
  \ket{\gamma_{A}} \otimes \ket{\gamma_{B}} 
  = \left(
      \alpha_{A} \begin{bmatrix}1 \\ 0\end{bmatrix} 
      + \beta_{A} \begin{bmatrix}0 \\ 1\end{bmatrix} 
    \right) 
    \otimes
    \left(
      \alpha_{B} \begin{bmatrix}1 \\ 0\end{bmatrix} 
      + \beta_{B} \begin{bmatrix}0 \\ 1\end{bmatrix} 
    \right)
  = \begin{bmatrix}
      \alpha_{A}\alpha_{B}\\
      \alpha_{A}\beta_{B}\\
      \beta_{A}\alpha_{B}\\
      \beta_{A}\beta_{B}\\
    \end{bmatrix}
  = \begin{bmatrix}
      p(s_{A}\land s_{B})^{\frac{1}{2}} \\
      p(s_{A}\land s_{B}^{\perp})^{\frac{1}{2}} \\
      p(s_{A}\land s_{B})^{\frac{1}{2}} \\
      p(s_{A}^{\perp}\land s_{B}^{\perp})^{\frac{1}{2}} \\
    \end{bmatrix}
    \begin{matrix}
      (a) \\
      (b) \\
      (c) \\
      (d) 
    \end{matrix}
}
where $ p(s) $ is the probability of state $ s $ occuring.
Note that $ s_{A} $ and $ s_{B} $ are distinct propositions, and if both pairs (a) and (d) (or (b) and (c)) occur with high probability, 
subsystems $ \gamma_{A} $ and $ \gamma_{B} $ are in "\emph{correlated states}". 

When some distinct propositional elements are correlated, 
there exist states in the system that can be called \emph{correlated states}. 
Redefining them as a single state (or we might be able to say a "concept") helps to decipher complex systems 
by making the relationship between time-varying states clearer.
Analogically, when each state refers to each finger of the right hand being stretched, 
the event-state system is decoded by redefining the coupled state as "the right hand is open”. \\
Let the bases corresponding to $ s_{p} $ and $ s_{q} $ be $ \ket{p} $, $ \ket{q} $.
The bases $ \ket{p} $, $ \ket{q} $ are orthogonal to each other on these spanned minimal Hilbert spaces $ \pazocal{H}^{4} $, 
but the basis $ \ket{p \land q} $ by linear combination of $ \ket{p} $, $ \ket{q} $ 
is no longer orthogonal to the original $ \ket{p} $, $ \ket{q} $ (Figure 1).
%\dmath{
%  (figure)
%}
In non-distributive logic on orthomodular lattices, 
two propositions $p$, $q$ are compatible if and only if
\eq{}{
  p=(p \land q) \lor (p \land q')\ and\ q = (q \land p) \lor (q \land p') 
}
Conversely, if $p$ and $q$ are incompatible, 
it means that $(p' \land q)$ and $(p' \land q')$ are true even if $p$ is true.
\begin{figure}[!t]%
  \centering
  \includegraphics[width=1.0\textwidth]{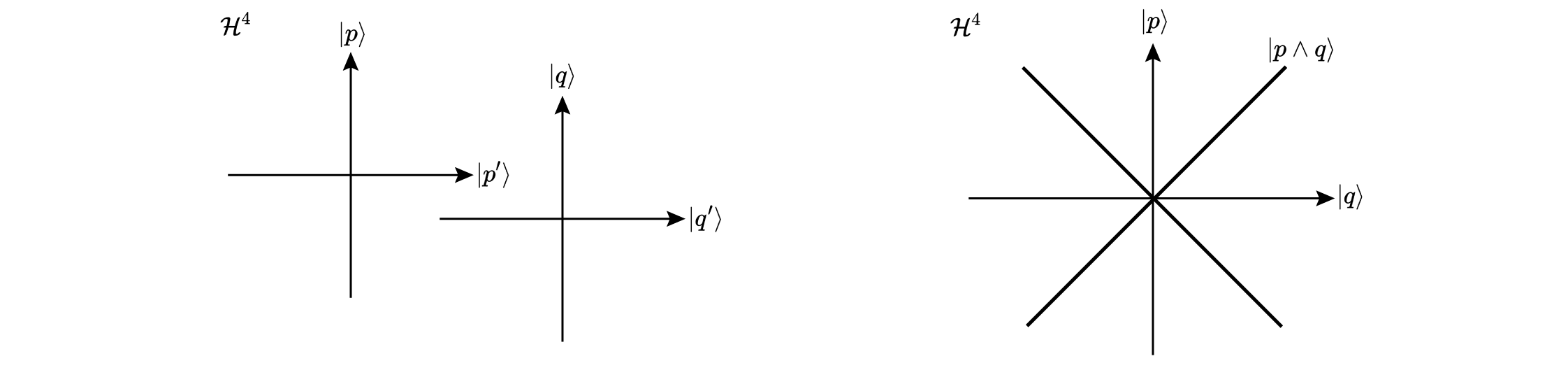}
  \caption{
    $p$ and $q$ are orthogonal on $ \hilbert^{4} $ , while $p$(or $q$) and 
    $ p \land q $ are not orthogonal (i.e., $p$ and $q$ are compatible, 
    while $p$(or $q$) and $ p \land q $ are incompatible).
    $ \hilbert^{4} $ is the simplest non-trivial composite system that can represent correlation information.
    Note that $\ket{p}$, $\ket{q}$,$\ket{p'}$, $\ket{q'}$ are orthogonal to each other in both left and right figures.
    }\label{fig1}
\end{figure}
As mentioned in Chapter 1,
in the structure of non-distributive quantum logic on orthogonal modular lattices, 
$ (S_{1} \land S_{2} \land \cdots \land S_{R} ) $ can be true even if 
$ S_{1},\ S_{2}, \ldots ,\ S_{R} $ are all false.
The computational language $\beta$ provides a formalization for manipulating this logical structure, as shown below.

Here, I take a subset $ \pazocal{N}' \subset \pazocal{N} $ 
and represent its linear combination by 
\eq{2.7}{
  \ket{\Phi} = \sum_{m \in \mathbb{Z}_{\lor}}\alpha_{m}\ket{m}\ \ (\mathbb{Z}_{\lor} = \{m\ |\ m\ \text{is\ an\ index\ of\ basis\ states\ in\ }\pazocal{{N}'}\})
}
where $ \mathbb{Z}\lor $ is the set of indices of the states in $ \pazocal{N}' $.
In quantum theory, $ \ket{\Phi} $ is a pure state when it is contained in the eigenspace 
$ E(A, \lambda) = \{\mathbfit{v} \in \hilbert^{\bar{M}}\ and\ A\mathbfit{v}=\lambda\mathbfit{v} \} $
of the observable (Hermitian operator), 
and constitutes a closed $1$-dimensional subspace on the subspace spanned by $ \pazocal{N}' $, 
which corresponds to $ \ket{p \land q} $ as mentioned above.
To clearly distinguish this newly generated \emph{combined state} from stochastic mixed states, 
in which it is simply uncertain which of $ \{ s_{m} | m \in z_{\lor} \} $ will occur, 
Let us denote it by $ \bigwedge_{m \in z_{\lor}} s_{m} $. 
The combined states are assigned to a different variable memory space on $ \ket{\xi} $ from the original states.
Thus the semantics of combined and mixed states are
\\
\\
\indent 1) if $ \ket{\Phi} $ is a \emph{combined state}, 
then $ \bigwedge_{m \in z_{\lor}} s_{m} $ is true (false) 
irrespective of $ s_{m} (m \in z_{\lor}) $
\\
\\
\indent 2) if $ \ket{\Phi} $ is a \emph{mixed state}, 
then $ \exists m \in z_{\lor} $, $ V_{m \in z_{v}}  s_{m} $ is true.
\\
\\
In particular, in 1), two propositions that are equivalent in classical logic are mapped to 
different closed subspaces on $ \mathbb{L}(\hilbert) $. 
Thus, the combined states connected by a conjunction are evaluated independently of the truth or falsehood 
of the individual propositions that form the subset.
An example of program codes that represent this logic is
\\
\\
\\
\newcommand{\code}[1]{\texttt{#1}}
\code{
    %{\setlength{\parindent}{0cm}
    1 \indent if any($ s_{m} $) \ \textit{\% any event in $ \{ s_{m} \} $}\\
    %}
    2 \indent \indent	print(`at least one event of $ s_{p}, s_{q}, s_{r}, \ldots $ is true') \\
    3 \indent elif  $ s_{m}^{\ast} $  \  \textit{\% combined events  $ \bigwedge_{m \in z_{\lor}} s_{m} $ } \\
    4 \indent \indent print(`$ s_{p} \land s_{q} \land s_{r}, \ldots $ is true') \\
}
\\
Note that the fourth line is never executed in a conventional framework.\\
\indent Let us assume, for example, that we define the basis state $ \mathbfit{c}_{1} $ 
of the minimal propositional element on some $ \hilbert^{2} $ along the eigenvectors of any operator 
belonging to the class $ \Omega $ defined in (1.17). 
Then, another combined state $ \mathbfit{c}_{2} $ is generated by a unitary transformation 
with $ \mathbfit{c}_{1} $ as the ground basis.
This recursive operation can generate an infinite number of basis states through repeated rotations 
involving irrational numbers. 
As a result, an infinite number of different $1$-dimensional subspace propositions 
can be generated in the plane, which never coincide.

Let Hermitian operators that output the eigenvalue $ m $ for the \emph{standard state} $ s_m \ (m \in \mathbb{Z}_{\bar{M}}) $ be
\eq{2.8}{
  A = \sum_{m \in \mathbb{Z}_{\lor}} m \ket{m} \bra{m}
}
then, by the Spectrum Theorem, observables that measure the combined state $ \ket{\Phi} $ can be written as
\eq{2.9}{
  A^{\ast} = \sum_{m \in \mathbb{Z}_{\lor}} m^{\ast} \ket{m^{\ast}} \bra{m^{\ast}}
  + \sum_{m \in \mathbb{Z}_{\lor}^{C}} m \ket{m}\bra{m}
}
Let $ \pazocal{N}' \subset \pazocal{N} $ be the basis states selected to generate the combined state $ \ket{\Phi} $, 
and $ \mathbb{Z}_{\lor} $ be its set of indices. 
$ \mathbb{Z}_{\lor}^{C} $ then represents the complement of $ \mathbb{Z}_{\lor} $ within $ \mathbb{Z}_{\bar{M}} $. 
Also let
\eq{2.10}{
  m^{\ast} = M + \prod_{m \in \mathbb{Z}_{\lor}} e_{m}\ \ 
  (e_{m}\ \text{is\ the\ m\ th\ prime\ number})
}
be the index of the newly generated $ |\mathbb{Z}_{\lor}| $ orthonormal bases.\\
New bases including $ \ket{\Phi} $ can be obtained by applying the Gram-Shmidt process
\eq{2.11}{
g(\Phi,\ \Phi',\ \cdots) 
  = \ket{\Phi} 
  - \left( \cdots 
             + \frac{\braket{\Phi''|\Phi}}{\braket{\Phi''|\Phi''}}\ket{\Phi''}
             + \frac{\braket{\Phi'|\Phi}}{\braket{\Phi'|\Phi'}}\ket{\Phi'}
    \right)\ \
    (\text{\'\ denotes\ the\ number\ of\ recursions})
}
recursively $ |\mathbb{Z_{\lor}}|-1 $ times. 
Let the linear combination of these newly generated bases be
\eq{2.12}{
\ket{\Phi} = \sum_{m^{\ast} \in \mathbb{Z}_{\lor}} \alpha_{m^{\ast}} \ket{m^{\ast}}
}
A unique index can then be output through the norm $ m^{*} $ of $ A^{\ast} \ket{\Phi} $.

When a combined state $ \Phi $ is generated, an orthogonal basis to $ \Phi $ is also automatically defined in the space $ \mathbb{L}(\hilbert)$.
In the variable memory space corresponding to this basis, 
a complementary proposition that is always false when $ \Phi $ is true is assigned along with the variable definition. 
These orthogonal propositions can be called "\textit{complementary pairs}" 
and are assigned a special type in the formalization of $ \beta $.

$\pazocal{C}$ is a mathematical model for massively parallel computing that defines combined or correlated states 
as logical propositions of a composite probability distribution, 
and enables programmatic manipulation of these states, which are not intuitively accessible to us. 
$\pazocal{C}$ is designed to determine a class $ <E,\,S> $ of event-state systems through large-scale linear properties, 
while finding nonlocal correlations contained in the complex composite system.
$\pazocal{C}$ and its computational language, $\beta$, 
deduce a generation or transformation process from a finite number of possible combinations of states at a given timestep $ t $ in parallel. 
For this purpose, the computational basis state $ \ket{\psi}(t) $ at each time step is manipulated by \emph{programs} 
$ \pazocal{P}: \ket{\psi}(t) \rightarrow \ket{\psi}(t+1) $, which are called \emph{hypotheses}. 
And spectra 
\eq{2.14}{
  \ket{\phi}(t) = \{\phi_{i} \}\ \ (i \in \mathbb{N}_{0}^{2\bar{M}-1}) \ \ \ \ \ \  \\
%}
%\dmath{
  \begin{cases}
    \phi_{2m-2} : \ \text{a\ frequency\ of\ states\ } s_{m} \\
    \phi_{2m-1} : \ \text{a\ frequency\ of\ states\ } s_{m}^{\perp}
  \end{cases}
}
of joint probability distributions are output through multiple projection measurements. 
These spectra are then separated into mutually exclusive states $ \ket{\phi}(t) \rightarrow \{ \ket{\phi}(t, k) \} $
and stored in memory $ \ket{\xi} $ to serve as input values to the instruction register at the next time step.
The Von-Neumann entropy $ \pazocal{S} $ of density operators are computed from the normalized joint probability distribution $ \ket{\phi}'(t) $ 
and the set of bases $ \pazocal{N} $. 
When the entropy becomes $ \pazocal{S} < \epsilon $, 
the class $ <E,\,S> $ of the event-state system is "\emph{decided}" and the system is said to have "\emph{converged}". 
The number of steps $ T $ required for the hypothesis to make the system converge can be regarded 
as the \emph{logical depth} of the system.

\subsection{Turing Completeness of $\pazocal{C}$}

\indent 
The instruction register  
$ g(\mathbfit{i},\mathbfit{x},\mathbfit{y},\phi(t),k) $
described in (\ref{eq:register})
is equivalent to a primitive recursive function
\eq{2.2.2}{
\begin{cases}
    h(\ket{\phi},\ 0) = f(\ket{\phi})  \\
    h(\ket{\phi},\ k+1) = g(\ket{\phi},\ k,\ h(\ket{\phi}, k))
\end{cases}
}
Furthermore, given a function 
\eq{2.2.3}{
  g: \hilbert \in \mathbb{C}^{M} \rightarrow \mathbb{N} \ \ \ \  \\
  f: \mathbb{N} \rightarrow \mathbb{N} 
}
we can define a function 
\eq{2.2.4}{
  f^{\ast} = gfg^{-1} \ \text{and} \  f^{\ast}: \mathbb{N} \rightarrow \mathbb{N}
}
that maps any transformation on $ \hilbert \in \mathbb{C}^{N} $ to $ \mathbb{N} \rightarrow \mathbb{N} $.
Thus, a set of primitive operators that can compute any point on $ \hilbert $ creates a class of $\mu$-recursive functions. 
Such a set of operators is trivially definable (e.g., invertible operators in $N$-dimensional Hilbert space which gives arbitrary rotation and expansion), 
and thus $ \pazocal{C} $ is Turing-complete.

The important problem is that the operator giving the transformation of any 
$ \mathbb{N} \ \text{(time\ step\ t)} \rightarrow \mathbb{N}  \ \text{(time\ step\ t+1)} $ is nontrivial. 
For example, it is not easy to deduce the existence of an operator $ \pazocal{F} $, 
just from the generated result $ \mathbb{N} \ \text{(time\ step\ t >\!> 0)}$, 
which gives the convergence of a point sequence on $ \hilbert $ of the formula (\ref{eq:fibmatrix}). 
Determining how the system is reasonably transformed is the primary motivation of $ \pazocal{C} $ and $ \beta $.

\subsection{Conditional Operators to Joint Probability Distributions}

\indent 
The hypothetical program to verify the transformation $ \ket{\psi}(t) \rightarrow \ket{\psi}(t+1) $ proceeds through a recursive execution 
that consists mainly of a conditional expression that takes a joint probability distribution $ \ket{\psi}(t) $ at any time $ t $ as an argument, 
and a transformation that updates any state of the subsystem.
%}
Conditional expressions evaluate the boolean values of variables defined as the combined and 
correlated states as well as the standard state, 
and are expressed in terms of input values to the instruction register $ g(\mathbfit{i},\mathbfit{x},\mathbfit{y},\phi(t),k) $.
The operators that realize these generalized representations
\eq{conditionalgeneral}{
  \text{if}\ \bigvee_{m \in \pi(\mathbfit{x})} s_{m} \ \text{is\ true\ (false),\ perform} \ X \  \text{to} \ \mathbfit{y}
}
in $\beta$ are discussed below.\\
Here, $ \bigvee_{m \in \pi(\mathbfit{x})} s_{m} $ is any standard state pair $ s_{u}\lor s_{v}\lor s_{w}\lor \cdots $, and $ \mathbfit{x} $ is the source states input to 
$ g(\mathbfit{i},\mathbfit{x},\mathbfit{y},\phi(t),k) $.
The source states are specified by $ \mathbb{Z}_{2} $, and hence
\eq{2.3.1}{
  \pi(\mathbfit{x}) = \{ m \in \bar{M} \ | \ x_{m} \in \mathbfit{x}  \ and \ x_{m} = 1  \}
}
Firstly, a subclass of (\ref{eq:conditionalgeneral}) 
\eq{conditionalsub}{
  \text{if}\ s_{m} \ \text{is\ true\ (false),\ perform} \ X \  \text{to} \ \mathbfit{y}
}
is realized as follows.
Since $ \ket{\phi} = \{ \phi_{i} \} \ (i \in \mathbb{N}_{0}^{2\bar{M}-1}) $ 
outputs the spectrum of the joint probability distribution of $ s_{m} $ for $ i = 2m-1 $ and $ s_{m}^{\perp} $ for $ i = 2m-1 $, 
the $ \bar{M} \times 2 \bar{M} $ operator can be thought as follows:
\eq{2.3.1}{
  \mathbfit{V}(\ket{\phi}) =
  \begin{bmatrix}
    1 & 0 & \cdots &     &         &        &        &  \\
      & 0 & 1      & 0   & \cdots  &        &        &  \\
      &   & 0      & 0   & 1       & \cdots &        &  \\
      &   &        &     & 0       & 0      & \cdots &  \\
      &   &        &     &         &        & \ddots & 
  \end{bmatrix}
}
\eq{2.3.1}{
  \mathbfit{V'}(\ket{\phi}) =
  \begin{bmatrix}
    0 & 1 & \cdots &     &         &        &        &  \\
      & 0 & 0      & 1   & \cdots  &        &        &  \\
      &   &        & 0   & 0       & 1      & \cdots &  \\
      &   &        &     &         & 0      & 0      &  \\
      &   &        &     &         &        & \ddots & 
  \end{bmatrix}
}
such that the matrix element $ a_{ij} $ is 
$ \delta^{(i,\ j)}_{(\forall m,\ 2m-1)} \ (m \in \bar{M})$\ if "\textit{is true}" and 
$ \delta^{(i,\ j)}_{(\forall m,\ 2m-2)} \ (m \in \bar{M})$\ if "\textit{is false}".
Thereby, as in
\eq{2.3.1}{
  \mathbfit{V}(\ket{\phi}) = 
  \begin{bmatrix}
    s_{1} \\
    s_{2} \\
    \vdots \\
    s_{\bar{M}} \\
  \end{bmatrix}
  ,\
  \mathbfit{V'}(\ket{\phi}) = 
  \begin{bmatrix}
    s_{1}^{\perp} \\
    s_{2}^{\perp} \\
    \vdots \\
    s_{\bar{M}}^{\perp} \\
  \end{bmatrix}
}
the output organized into $ s_{m} $ and $ s_{m}^{\perp} $ is
\eq{2.3.1}{
  \ket{\phi'}=\{ \phi'_{i} \} \ \ (i \in \mathbb{\bar{M}}-1)
}

Then the operator $ \mathbfit{IF}^{x}_{\mathbf{y}} $, 
which outputs $ \phi_{2m-2} $ (or $ \phi_{2m-1}$) $ \in \mathbb{N} $ 
for the specified target state $ \mathbfit{y} $ 
only if the specified source state $ x $ is true (or false), 
and $ 0 $ for all others, 
is an $ \bar{M} \times \bar{M} $ matrix such that the element $ a_{i,\ j} $ is 
$ \{ \delta^{\phi'_{x} y_{i} }_1 \} \ (i \in \mathbfit{y})$
given by 
\eq{IFV}{
  \mathbfit{IF}^{\ast}
  \mathbfit{V}^{(')}(\ket{\phi}) = 
  \begin{bmatrix}
    0      & \cdots & \delta^{\phi'_{x} y_{1} }_1       & \cdots & 0 \\
    0      & \cdots & \delta^{\phi'_{x} y_{2} }_1       & \cdots & 0 \\
    \vdots &        & \vdots                            &        & \vdots \\
    0      & \cdots & \delta^{\phi'_{x} y_{m} }_1       & \cdots & 0 \\
    \vdots &        & \vdots                            &        & \vdots \\
    0      & \cdots & \delta^{\phi'_{x} y_{\bar{M}} }_1 & \cdots & 0 \\
  \end{bmatrix}
  \mathbfit{V}^{(')}(\ket{\phi})
  =
  \begin{bmatrix}
    \delta^{\phi'_{x} y_{1} }_1 \phi_{1}' \\
    \vdots \\
    \delta^{\phi'_{x} y_{\bar{M}} }_1 \phi_{\bar{M}}' 
  \end{bmatrix}
}
where $\mathbfit{V}^{(')} $ represents $ \mathbfit{V} $ or $ \mathbfit{V}' $.
To generate the conditional expression given in (\ref{eq:conditionalgeneral}) as an extension to the super class of (\ref{eq:conditionalsub}),
%\dmath{
%  \text{if}\ \bigvee_{m \in \pi(\mathbfit{x})} s_{m} \ \text{is\ true\ (false),\ perform} \ X \  \text{to} \ \mathbfit{y}
%}
we only need to take the direct sum of 
$ \mathbfit{IF}^{x_{m}}_{\mathbfit{y}} \mathbfit{V}^{(')}(\ket{\phi}) $ for all specified $ x_{m} \in \mathbfit{x} $.
Therefore, this can be written as
\eq{2.3.1}{
  \mathbfit{IF}^{x_{u}}_{\mathbfit{y}} \mathbfit{V}^{(')}(\ket{\phi})
  \oplus
  \mathbfit{IF}^{x_{v}}_{\mathbfit{y}} \mathbfit{V}^{(')}(\ket{\phi})
  \oplus
  \cdots
  \mathbfit{IF}^{x_{z}}_{\mathbfit{y}} \mathbfit{V}^{(')}(\ket{\phi})
  =
  \bigoplus_{m \in \pi({\mathbfit{x}})}
  \mathbfit{IF}^{x_{m}}_{\mathbfit{y}} \mathbfit{V}^{(')}(\ket{\phi})
}
Regarding logical product
\eq{2.3.1}{
  \text{if}\ \bigwedge_{m \in \pi(\mathbfit{x})} s_{m} \ \text{is\ true\ (false),\ perform} \ X \  \text{to} \ \mathbfit{y}
}
similarly the output $ \{ \delta^{\phi'_{x} y_{m} }_1 \phi_{\bar{m}}' \} \ (m \in \bar{M}) $ 
can be obtained as in (\ref{eq:IFV}) by taking the Adamard product of
$ \mathbfit{IF}^{x_{m}}_{\mathbfit{y}} \mathbfit{V}^{(')}(\ket{\phi}) \ (m \in \pi(\mathbfit{x}))  $
, i.e.,
\eq{2.3.1}{
  \mathbfit{IF}^{x_{u}}_{\mathbfit{y}} \mathbfit{V}^{(')}(\ket{\phi})
  \odot
  \mathbfit{IF}^{x_{v}}_{\mathbfit{y}} \mathbfit{V}^{(')}(\ket{\phi})
  \odot
  \cdots
  \mathbfit{IF}^{x_{z}}_{\mathbfit{y}} \mathbfit{V}^{(')}(\ket{\phi})
  =
  \bigodot_{m \in \pi({\mathbfit{x}})}
  \mathbfit{IF}^{x_{m}}_{\mathbfit{y}} \mathbfit{V}^{(')}(\ket{\phi})
}

Furthermore, the conditional expression by any combination of logical 
ORs $ \text{if}\ \bigvee_{m \in \pi(\mathbfit{x})} s_{m} $ 
and 
ANDs $ \text{if}\ \bigwedge_{m \in \pi(\mathbfit{x})} s_{m} $ 
can be constructed from any concatenation by $ \oplus $ and $ \odot $ for 
$ \mathbfit{IF}^{x_{m}}_{\mathbfit{y}} \mathbfit{V}^{(')}(\ket{\phi}) \ (m \in \pi(\mathbfit{x}))  $
.

\section{Conclusions and Further Research}\label{sec3}

To decodeh highly complex symbol sequences generated by systems with computational universality,
%When a symbol sequence with high complexity is generated by a system with computational universality, 
a general computational method is proposed that derives a transformation process (generation rule) 
from the generation results.
This transformation process gives complexity or logical depth in terms of computation time, which efficiently provides
$ \mathbb{Z} \rightarrow \mathbb{Z} $ generation.
I formulated a universal diductive machine model $ \pazocal{C} $ and 
a computational language $ \beta $ that relies on the structure of non-distributive quantum logic 
$ \mathbb{L}(\hilbert) $ in Hilbert spaces or orthogonal modular lattices, 
i.e., a non-boolean logic structure in which no proposition implying a generated outcome 
can be uniquely determined. 
The $ \pazocal{C} $ and its language $ \beta $ performs massively parallel computations on event-state systems consisting of 
exponential combinations of propositional elements, 
which are represented by correlations between subsystems. 
To construct and operate this basic framework, a model of the theory of computation is presented, 
as well as the conditional logical operations performed on the framework among its other main operations.
At the same time, as a generalization of functions that show evidence of a long evolutionary history, 
a class of operators with quasi-periodic structures whose eigenvectors (or eigenvalues) include 
irrational numbers as components is defined as $ \Omega $.

As for future prospects, first of all, theoretically, it will be necessary to study pure mathematics, 
such as the theory of rings of operators for the symmetry of transformations or something like 
surreal numbers with non-classical algebraic rules of numbers. 
Second, from an empirical point of view, 
detailed observations of real (terrestrial) natural systems will be needed, 
especially observations of communication between systems that perform life activities 
in the broadest sense of the word. 
Third, it will be necessary to develop a computational language and computer itself 
to substantiate or simulate the model.
\\

This work has established a foothold, 
but it is only the beginning, 
and the road ahead is endless. 
Is what lies beyond that something like what Mandelbrot (2012) calls
\textit{"the ultimate knowledge, likened to the Holy Grail or the Golden Fleece"}?
No one knows. 
The question raised at the beginning of this paper is one that I have been asking myself since I was a teenager. 
I would like to thank all those who have supported and given me the opportunity to do this research.

  %USE THE BELOW OPTIONS IN CASE YOU NEED AUTHOR YEAR FORMAT.
  %\bibliographystyle{abbrvnat}
  %\bibliography{reference}

\begin{thebibliography}{10}

  % Anderson, Philip W. 
  % "More is different: broken symmetry and the nature of the hierarchical structure of science." 
  % Science 177.4047 (1972): 393-396.
  \bibitem{anderson1972}
  Anderson, Philip W.
  \newblock More is different: broken symmetry and the nature of the hierarchical structure of science.
  \newblock {\em Science}, 177.4047: 393-396., 1972.

  % Baake, Michael. 
  % “A Guide to Mathematical Quasicrystals.” 
  % arXiv: Mathematical Physics (1999): 17-48.
  % \bibitem{baake1999}
  % Baake, Michael. 
  % \newblock A Guide to Mathematical Quasicrystals.
  % \newblock {\em arXiv: Mathematical Physics}, 17-48, 1999.
 
  % Bennett, C. H. (1988). Logical depth and physical complexity. In R. Herken (ed.), The universal Turing machine, a half century survey. Oxford University Press. pp. 227-257.
  \bibitem{benett1988}
  Bennett, C. H.
  \newblock Logical depth and physical complexity. In R. Herken (ed.), The universal Turing machine, a half century survey.
  \newblock {\em Oxford University Press.}, pp. 227-257., 1988.

  % Bennett, Charles H. 
  % "Complexity in the universe." 
  % Physical origins of time asymmetry (1994): 33-46.
  \bibitem{benett1994}
  Bennett, C. H.
  \newblock Complexity in the universe.
  \newblock {\em Physical origins of time asymmetry}, 33-46., 1994.

  % Bennett, Charles & Gacs, Peter & Li, Ming & Vitányi, Paul & Żurek, Wojciech. (1998). 
  % Information Distance. Information Theory, 
  % IEEE Transactions on. 44. 1407 - 1423. 10.1109/18.681318.
  \bibitem{benett1998}
  Bennett, Charles \and Gacs, Peter \and Li, Ming \and Vit\'anyi, Paul \and \.Zurek, Wojciech.
  \newblock Information Distance. Information Theory 
  \newblock {\em IEEE Transactions on. 44. 1407 - 1423. 10.1109/18.681318.}, 1998.
 
  % Birkhoff, G., and J. von Neumann, 
  % The logic of quantum mechanics. 
  % Ann. of Math. 37(1936), 823-843.
  \bibitem{birkhoffAndNeumann1936}
  Birkhoff, G., and J. von Neumann
  \newblock The logic of quantum mechanics.
  \newblock {\em Ann. of Math.}, 37(1936), 823-843.
    
  % Cohen, D.W. (1989) 
  % An Introduction to Hilbert Space and Quantum Logic. 
  % Springer-Verlag, 
  % 94-104. https://doi.org/10.1007/978-1-4613-8841-8
  \bibitem{cohen1989}
  Cohen, D.W.
  \newblock An Introduction to Hilbert Space and Quantum Logic.
  \newblock {\em Springer-Verlag}, 94-104. https://doi.org/10.1007/978-1-4613-8841-8, 1989.

  % Cook, Matthew (2004). 
  % "Universality in Elementary Cellular Automata" (PDF). 
  % Complex Systems. 15: 1–40.
  \bibitem{cook2004}
  Cook, Matthew
  \newblock Universality in Elementary Cellular Automata
  \newblock {\em Complex Systems. 15: 1–40.}, 2004.

  % Deutsch, D. (1985). 
  % Quantum theory, the Church–Turing principle and the universal quantum computer. 
  % Proceedings of the Royal Society of London. A. Mathematical and Physical Sciences, 400, 117 - 97.
  \bibitem{deutsch1985}
  Deutsch, D.
  \newblock Quantum theory, the Church–Turing principle and the universal quantum computer. 
  \newblock {\em Proceedings of the Royal Society of London. A. Mathematical and Physical Sciences}, 400, 117 - 97., 1985.
 
  % Deutsch, D. (1989) 
  % Quantum Computational Networks. 
  % Proceedings of the Royal Society of London A, 425, 73-90.
  % http://dx.doi.org/10.1098/rspa.1989.0099
  \bibitem{deutsch1989}
  Deutsch, D.
  \newblock Quantum Computational Networks.
  \newblock {\em Proceedings of the Royal Society of London A}, 425, 73-90.
  http://dx.doi.org/10.1098/rspa.1989.0099, 1989.

  % H. Bohr, 
  % "Almost-periodic functions", Chelsea, reprint (1947)
  \bibitem{bohr1947}
  H. Bohr
  \newblock Almost-periodic functions.
  \newblock {\em Chelsea, reprint}, 1947.

  % Hofstadter, D. R. J 1979 
  % Godel, Escher, Bach: an eternal golden braid. 
  % ¨ New York: Random House.
  \bibitem{hofstadter1979}
  Hofstadter, D. R. J
  \newblock Godel, Escher, Bach: an eternal golden braid.
  \newblock {\em New York: Random House.}, 1979.
 
  % Mandelbrot, Benoit (2012). 
  % The Fractalist: Memoir of a Scientific Maverick. 
  % Pantheon Books. ISBN 978-0-307-38991-6.
  \bibitem{mandelbrot2012}
  Mandelbrot, Benoit.
  \newblock The Fractalist: Memoir of a Scientific Maverick.
  \newblock {\em Pantheon Books. ISBN 978-0-307-38991-6.}, 2012.
 
  % Minsky, M. (1961). 
  % Recursive Unsolvability of Post's Problem of "Tag" and other Topics in Theory of Turing Machines. 
  % Annals of Mathematics, 74, 437.
  \bibitem{minsky1961}
  Minsky, M.
  \newblock Recursive Unsolvability of Post's Problem of "Tag" and other Topics in Theory of Turing Machines.
  \newblock {\em Annals of Mathematics}, 74, 437., 1961.

  % Mermin, N.D. 
  % The Ithaca interpretation of quantum mechanics. 
  % Pramana - J Phys 51, 549–565 (1998). https://doi.org/10.1007/BF02827447
  \bibitem{mermin1998}
  Mermin, N.D. 
  \newblock The Ithaca interpretation of quantum mechanics.
  \newblock {\em Pramana - J Phys 51}, 549–565. https://doi.org/10.1007/BF02827447, 1998.
 
  % Peres, Asher (1995). 
  % Quantum Theory: Concepts and Methods. Kluwer. 
  % ISBN 9780792336327. OCLC 901395752
  \bibitem{peres1995}
  Peres, Asher.
  \newblock Quantum Theory: Concepts and Methods. 
  \newblock {\em Kluwer. ISBN 9780792336327. OCLC 901395752}, 1995.
  
  % Putnam, H. (1969). 
  % Is Logic Empirical? 
  % Boston Studies in the Philosophy of Science, 
  % 216–241. doi:10.1007/978-94-010-3381-7_5
  \bibitem{putnam1969}
  Putnam, H.
  \newblock Is Logic Empirical?
  \newblock {\em Boston Studies in the Philosophy of Science}, 216–241. doi:10.1007/978-94-010-3381-7-5, 1969.

  % Shannon, Claude E. 
  % "A universal Turing machine with two internal states." 
  % Automata studies 34 (1956): 157-165.
  \bibitem{shannon1956}
  Shannon, Claude E.
  \newblock A universal Turing machine with two internal states.
  \newblock {\em Automata studies 34}, 157-165., 1956.

  % von Neumann, J. (1932) 
  % Mathematische Grundlagen der Quantenmechanik. 
  % Springer, Berlin. English Translation by Beyer, R.T. (1955) 
  % Mathematical Foundations of Quantum Mechanics. Princeton University Press, Princeton.
  \bibitem{neumann1932}
  von Neumann, J.
  \newblock Mathematische Grundlagen der Quantenmechanik. 
  Springer, Berlin. English Translation by Beyer, R.T. (1955) Mathematical Foundations of Quantum Mechanics.
  \newblock {\em Princeton University Press, Princeton.}, 1932.


  \end{thebibliography}

\end{document}